\documentclass[pdflatex,mathphys]{sty}% Math and Physical Sciences Reference Style

\usepackage{upgreek}
\jyear{2022}%

\unnumbered% uncomment this for unnumbered level heads

\begin{document}

\title[Integrated photonic processor for broadband blind source separation]{Broadband physical layer cognitive radio with an integrated photonic processor for blind source separation}

\author*[1]{\fnm{Weipeng} \sur{Zhang}}\email{weipengz@princeton.edu}
\author[2]{\fnm{Alexander} \sur{Tait}}\email{alex.tait@queensu.ca}
\author[3]{\fnm{Chaoran} \sur{Huang}}\email{crhuang@ee.cuhk.edu.hk}
\author[4,1]{\fnm{Thomas} \sur{Ferreira de Lima}}\email{thomas@nec-labs.com}
\author[1]{\fnm{Simon} \sur{Bilodeau}}\email{sbilodeau@princeton.edu}
\author[1]{\fnm{Eric} \sur{Blow}}\email{ecb10@princeton.edu}
\author[1]{\fnm{Aashu} \sur{Jha}}\email{aashuj@princeton.edu}
\author[5]{\fnm{Bhavin J.} \sur{Shastri}}\email{bhavin.shastri@queensu.ca}
\author*[1]{\fnm{Paul} \sur{Prucnal}}\email{prucnal@princeton.edu}

\affil*[1]{\orgdiv{Department of Electrical and Computer Engineering}, \orgname{Princeton University}, \orgaddress{\city{Princeton}, \postcode{08540}, \state{New Jersey}, \country{USA}}}

\affil[2]{\orgdiv{Department of Electrical and Computer Engineering}, \orgname{Queen’s University}, \orgaddress{\city{Kingston}, \postcode{K7L 3N6}, \state{Ontario}, \country{Canada}}}

\affil[3]{\orgdiv{Department of Electronic Engineering}, \orgname{The Chinese University of Hong Kong}, \orgaddress{\city{Hong Kong}, \country{China}}}

\affil[4]{\orgname{NEC Laboratories America}, \orgaddress{\city{Princeton}, \postcode{08540}, \state{New Jersey}, \country{USA}}}

\affil[5]{\orgdiv{Department of Physics, Engineering Physics and Astronomy}, \orgname{Queen’s University}, \orgaddress{\city{Kingston}, \postcode{K7L 3N6}, \state{Ontario}, \country{Canada}}}

\abstract{The expansion of telecommunications incurs increasingly severe crosstalk and interference, and a physical layer cognitive method, called blind source separation (BSS), can effectively address these issues. BSS requires minimal prior knowledge to recover signals from their mixtures, agnostic to carrier frequency, signal format, and channel conditions. However, previous electronic implementations of BSS did not fulfill this versatility requirement due to the inherently narrow bandwidth of radio-frequency (RF) components, the high energy consumption of digital signal processors (DSP), and their shared weaknesses of low scalability. Here, we report a photonic BSS approach that inherits the advantages of optical devices and can fully fulfill its ``blindness" aspect. Using a microring weight bank integrated on a photonic chip, we demonstrate energy-efficient, WDM-scalable BSS across 19.2 GHz of bandwidth, covering many standard frequency bands. Our system also has a high (9-bit) resolution for signal demixing thanks to a recently developed dithering control method, resulting in higher signal-to-interference ratios (SIR) even for ill-conditioned mixtures.}


\maketitle

% \section{Introduction}

\begin{figure}[ht]
\centering
\includegraphics[width=.99\linewidth]{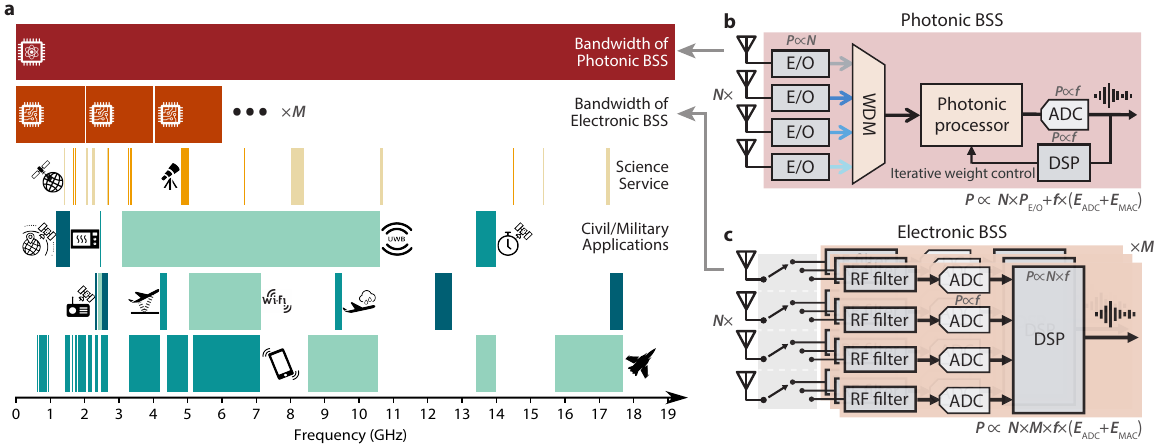}
\centering
\caption{\textbf{Architecture comparison between electronic and photonic BSS implementation.} \textbf{a}, Frequency allocation for common bands (from 0.5 GHz to 19 GHz) and example bandwidths for the two BSS implementations (photonics and electronics). Science service bands comprise earth explorer satellite service (EESS) and radio astronomy service (RAS). Bands for civil and military applications include global positioning system (GPS), microwave oven, ultra-wideband (UWB), and standard frequency and time signal service (for the first row); broadcasting-satellite service, radar altimeter, Wi-Fi, and aircraft weather radar (for the second row); 5G cellular networks and military radar (for the third row). \textbf{b}, Photonic BSS system diagram. \textbf{c}, Electronic BSS system diagram. In electronic BSS, covering the demanded bandwidth requires a complex switching system that consists of multiple electronic BSS setups (quantity of $M$) corresponding to distinct frequency regimes. Each one requires $N$ ADCs ($N$ is the number of receivers) and a dedicated DSP with $N$ inputs to perform the multiply-accumulate (MAC) operation. Since ADC and DSP require power in scale with the signal frequency, the total power consumption of the electronic BSS approaches is proportional to $N\times M \times f \times(E_\textrm{ADC}+E_\textrm{MAC})$. In contrast, achieving the same coverage requires only one photonic BSS setup consisting of N electrical-to-optical (E/O) converters with only one ADC and a low-end DSP with a single input. This architecture consumes much less power that is proportional to $N\times P_\textrm{E/O}+f(\times E_\textrm{ADC}+E_\textrm{MAC})$.
}
\label{fig:vision_figure}
\end{figure}

Many scientific activities, including the earth explorer services \cite{maeda2012eess} (for remote sensing, satellite imagery, radar) and radio astronomy \cite{burke2019ras}, must sensitively detect weak signals at frequencies naturally dictated by physical phenomena (Fig. \ref{fig:vision_figure}a). Thus, they are vulnerable to radio-frequency (RF) interference from commercial activities whose emission spectrum overlaps with the frequency range of interest. However, this vulnerability to RF interference is increasingly severe because of the expansion of telecommunications. As emerging wireless telecommunication signals \cite{navarro2020survey,dang2020_6g,6G_20} are squeezed into limited frequency bands, many strategies have been deployed to maximize spectrum utilization, such as the multi-input and multi-output (MIMO) scheme \cite{venkateswaran2010analog,larsson2014massive,kutty2015beamforming,han2015large,ghauch2016subspace}, which enhances data-carrying capacity through space-division multiplexing. The inevitable downside incurred is degraded signal quality due to severe crosstalk among tightly packed spatial channels. Also, when many wireless signals transmit simultaneously using frequencies that are close together, the raised spectral congestion negatively impacts the scientific services through RF interference.

One way to mitigate spectral congestion is active radio access sharing via cognitive radio \cite{haykin2005cognitive,akyildiz2006next,akyildiz2008survey}, which dynamically allocates secondary users with access to the unlicensed bands that do not have primary users. This method exploits many spectral gaps between bands licensed by regulatory bodies, which were skipped by the existing multiplexing schemes. However, this relies on a complicated software-layer radio-signal identification mechanism that is susceptible to privacy breaches \cite{clancy2008security,fragkiadakis2012survey}. Substantial signal processing and analysis are required to ascertain which transmissions are related to others and attributable to specific users. In situations with many users in wide frequency bands, performing these computations in real-time may not be possible. The only feasible way to oversee activity and enforce compliance from a regulatory standpoint may be to record spectral data for offline analysis, which introduces a serious risk to content privacy \cite{evfimievski2003limiting,acquisti2006there}. As soon as information is recorded to disk, its security is considered compromised. Even if the monitoring operator is deemed to be benign, it may be unknowingly harboring malware that can access the content of all the spectrum users.

A physical layer cognitive technique called blind source separation (BSS) \cite{choi2005blind, venkateswaran2010analog} can extract unknown signals (e.g., a signal of interest and an interferer) from their mixtures with minimal assumptions, as shown in Fig. \ref{fig:vision_figure}b and c. Operating in the physical layer allows the isolation of unwanted transmissions in the analog domain, which reduces the risk of privacy leakage by eliminating the need to record the transmission content digitally \cite{tait2018blind}. Another advantage of ``blindness'' is the agility in recovering sources with arbitrary characteristics, which means discarding a substantial amount of information prior to digitization and total obliviousness to the frequency, modulation type, and power ratio. This advantage can only be realized when BSS is performed across a wide frequency range. However, a broadband operation can be challenging for the electronic implementations of BSS due to the limited bandwidth of conventional RF technology. For example, the spectrum of ultra-wideband (UWB) signals \cite{UWB_04} covers up to 7.5 GHz, and that of Wi-Fi signals has expanded from 2.4 GHz (802.11) to 6 GHz (802.11ax). To have such broadband coverage is challenging with a single RF system, as depicted in Fig. \ref{fig:vision_figure}. Thus, alternative techniques other than conventional electronic processors are required to effectively process RF signals in the next generation of wireless systems \cite{lin2022artificial}.

By upconverting to frequencies of hundreds of terahertz, photonic signal processors can deal with broadband information \cite{marpaung2019integrated,shastri2021photonics,huang2021silicon}, where GHz signals are regarded as narrowband. As a result, photonic processing has low energy consumption that does not scale with the signal frequency. A promising on-chip processor is the microring resonator (MRR) weight bank \cite{mrr_16,de2022design}, a bank of tunable filters implemented with tiny circular optical waveguides, which provides energy-efficient tuning \cite{tait2022power} and scalable parallel processing through wavelength-division-multiplexing (WDM). Such an RF frontend powered by a photonic processor (Fig. \ref{fig:vision_figure}b) can share the workload of signal processing with a digital signal processing (DSP) backend and enhance the performance in many aspects. One key factor determining BSS performance is the resolution of the weights (the tuning accuracy of MRRs), which was reported up to 7 bits on MRRs \cite{mrrfb_18, huang2020demonstration}. With that said, we recently developed a dithering control method \cite{zhang2022silicon} that improves the tuning accuracy beyond 9 bits.

Here, we report a photonic implementation for BSS based on the dithering-controlled MRR weight bank. We also demonstrate a fully-packaged photonic processor with a silicon photonic chip integrated with MRR driver and control electronics on a single printed circuit board (PCB). We proved this setup could recover a weak transmitted signal in the presence of broadband jamming noise and successfully tested it in real-time on a wireless transceiver system. In terms of performance, our setup fully realizes the ``blindness" agility by achieving a bandwidth of up to 19.2 GHz and SIR of more than 40 dB in some cases. Besides, the dithering weight control enables a photonic processor with 9-bit accuracy beyond many electronic counterparts, enhancing the BSS with at least one-half reduced residual error than the setup without the dithering control. This work introduces the first functional BSS system capable of operating at broad bandwidths. When included in transceiver circuits, it can help cancel interference signals, with potential implications in cognitive radio and radio astronomy.

\section{BSS problem and algorithm}

The BSS problem can be formulated as separating an unknown mixture of unknown independent signals. The instantaneous model of generic mixing in a transmission channel is $[r_1,r_2] = \mathbf{H} [s_1, s_2]$, where $(s_1, s_2)$ are the source signals, $\mathbf{H}$ is a mixing matrix representing the wireless channel, and $(r_1, r_2)$ are the received signals. In the most general case, the source signals and their mixing matrix are unknown to the receiver. The goal of BSS is to estimate the corresponding demixing matrix, $\mathbf{H}^{-1}$, and apply it to the received signals to arrive at $[s'_1, s'_2] = \mathbf{H}^{-1}[r_1, r_2]$, where $(s'_1, s'_2)$ is the estimated recovery of the source signals. In addition to assuming nothing about the original signals, in general, minimal prior assumptions about their characteristics can be made. For example, $s_1$ and $s_2$ can occupy the same frequency band, meaning an implementation based on filtering would fail regardless of any receiver-side analysis. It is assumed that the two sources are statistically independent and that all mixing happened linearly. These assumptions are very realistic and are widely used in the radio community \cite{fabrizio2014blind}.

\begin{figure}[ht]
\centering
\includegraphics[width=.95\linewidth]{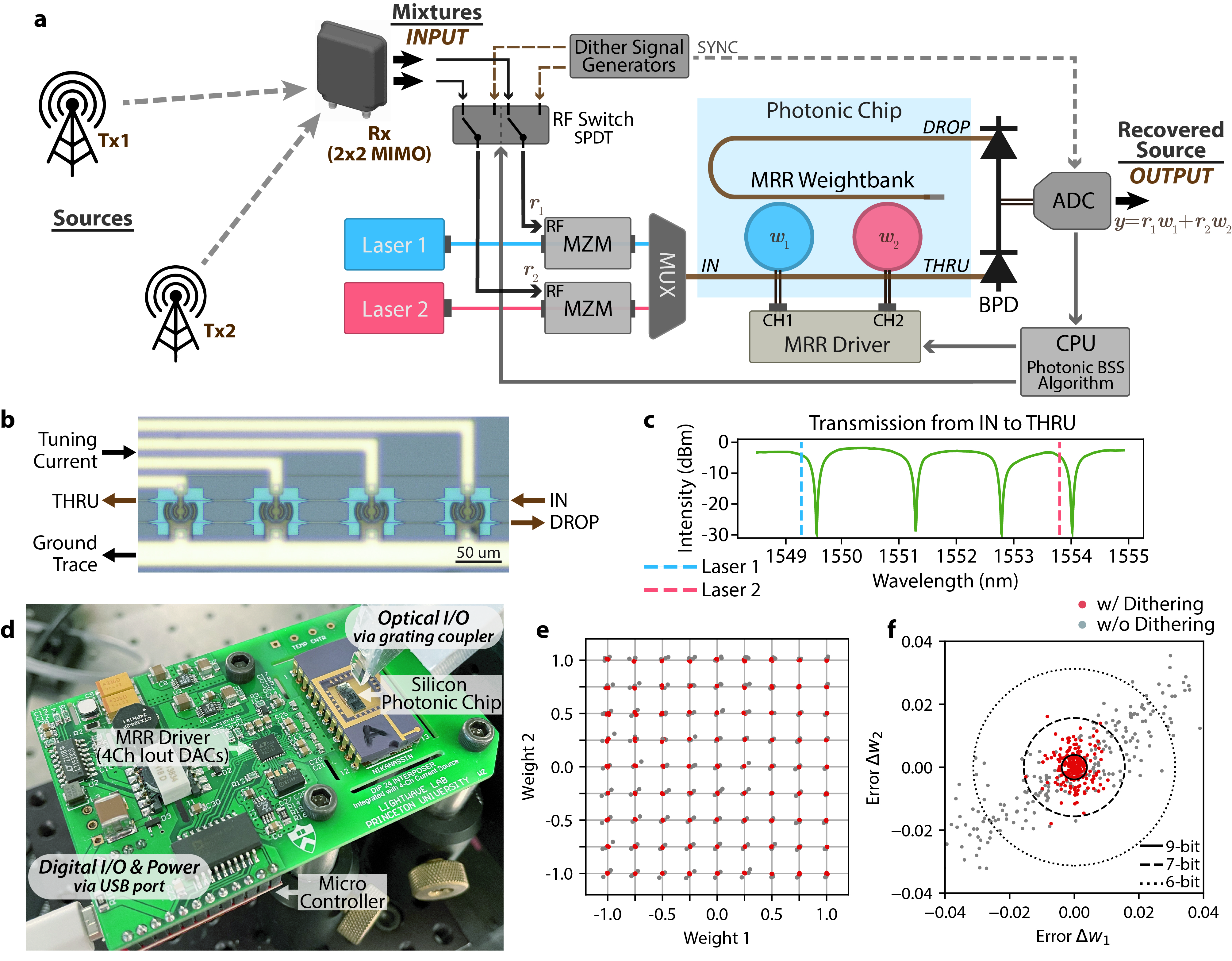}
\caption{\textbf{Photonic BSS experimental setup.} \textbf{a}, Schematic of the BSS setup. MZM, Mach-Zehnder modulator. MUX, wavelength-dependent multiplexer. BPD, balanced photodetector. ADC, analog to digital converter. Tx, transmitter. Rx, receiver. \textbf{b}, Micrograph of the MRR weight bank on the chip. \textbf{c}, Transmission spectrum of the THRU port at 25\textdegree C. \textbf{d}, Fully-packaged photonic processor. The silicon photonic chip, which is wire-bonded on a DIP 24 chip carrier, is mounted on the right side of this PCB. A 4-channel current output DAC is integrated on the broad and applies currents for tuning the MRRs. Optical input and output (I/O) is through the grating coupler on the top right, and electrical I/O is set up via a USB cable (bottom left), which also delivers the power for the whole PCB. \textbf{e}, Estimation of weighing accuracy. The 2-MRR weight bank in \textbf{a} was tested to tune the weights represented by each grid. The dithering control yielded the red points, and the gray points were obtained without the dithering. \textbf{f}, The errors of all the tested weights in \textbf{d}. A 9.0-bit of precision resulted from the dithering control and 6.7-bit for the control without the dithering.
}
\label{fig:schematic}
\end{figure}

For a given mixing matrix $\mathbf{H}$, to separate each signal component requires the mixtures to be weighted and summed with weights represented by each column of the inverse matrix $\mathbf{H^{-1}}$. This operation can be implemented with a set of two MRR weight banks acting as an on-chip signal processor. Retaining the signal of interest and eliminating the other one utilizes one of the two columns, which requires only one weight bank. As shown in Fig. \ref{fig:schematic}a-c, the MRR weight bank consists of several microring resonators with slightly different radii; each has a Lorentzian-shaped transmission profile (as shown in Fig. \ref{fig:schematic}c) centered at different wavelengths. Each MRR is equipped with a metal heater to allow thermo-optic displacement of the center wavelength by varying the current applied \cite{mrrcntr_16}. The MRR weight bank independently weights the laser amplitudes of different wavelengths. The sum of all optical powers can be obtained by a balanced photodetector (BPD) at the output. Utilizing this ability of weighted addition, we develop a photonic BSS algorithm, which follows a pipeline consisting of three steps, including principal component analysis (PCA) \cite{pca_19}, whitening, and independent component analysis (ICA) (See details in Ref. \cite{bss_20}). For carrying out these analyses, an iterative algorithm is preferred because of its simplicity in that only one vector needs to be commanded to the MRR weight bank in each step. Essentially, a constrained Nelder-Mead algorithm \cite{algm_00} is carried out that performs an iterative projection-pursuit of the mixtures to search the optimized weighting vectors. The goal is to find the mixture that outputs a weighted addition ($\Sigma w_i s_i, w_i \in [-1,1], i\in[1,2\cdots N]$, $N$ is the number of the mixtures) with maximal variance (the second-order statistic) for PCA and the maximal non-Gaussianity (the fourth-order statistic or kurtosis) for ICA.

\section{Photonic hardware implementation}

The hardware realization of this algorithm appears as a control loop (as shown in Fig. \ref{fig:schematic}a). Apart from the photonic chip, also included is a BPD for electrical-to-optical (E/O) conversion (DSC-R405ER, Discovery semiconductor), an ADC for signal digitization (DPO73304SX, Tektronix), a computer for statistical analysis and actuating weight commands, and a multi-channel current source for MRR tuning (custom-built as shown in Fig. \ref{fig:schematic}d). The dithering control \cite{zhang2022silicon} implemented here allows control of the MRRs with less complicated drivers instead of the source-measurement unit (SMU) \cite{mrrfb_18}. In this setup, the MRR driver is directly integrated into the PCB interposer, packaged close to the photonics chip with a much-reduced footprint and cable management \cite{zhang2022silicon}. The signal path starts from the MZM and ends at the scope, and the highest supported RF frequencies are determined by the BPD of up to 20 GHz, providing coverage for many commonly used RF bands. It is also worth noting that most of the signal path is in the optical domain, bringing about broadband and flat response and very low latency.

The photonic chip in this setup has a 4-channel MRR weight bank with resonance frequencies roughly spaced by 200 GHz. The spectra of the four MRRs (at 25 degrees) are shown in Fig. \ref{fig:schematic}c, with the resonance peaks located at 1549.6 nm, 1551.3 nm, 1552.8 nm, and 1554.0 nm. Since this work recovers source signals from two mixtures, we use two MRRs (the leftmost and rightmost ones). The corresponding lasers (PPCL500, Pure Photonics) are tuned to be 1549.3 nm and 1553.8 nm, then amplified (FA-23, Pritel) and combined into a shared waveguide by a WDM multiplexer (MDM-15-8, Santec) before coupling into the MRR weight bank through grating couplers.

The implemented dithering control method overcomes the low accuracy incurred by the high sensitivity of the weight bank. As shown in Fig. \ref{fig:schematic}a, the lasers are modulated with either the actual received mixtures or pre-defined dithering signals. Each time a set of commanded weights are applied, the RF switch (RC-2SPDT-A26, DC - 26.5 GHz, Mini-Circuits) passes the dithering signals into the photonic path, which helps adjust the driving currents of the MRRs until the output weights reach the demanded values. Then, the actual mixtures are switched into the weight bank and processed. The weight accuracy is reflected in error between the target and actual weights. Usually, we quantify the accuracy in bits, which is calculated as $\mathrm{log}_2(2/(w_\textrm{actual}-w_\textrm{target}))$. Fig. \ref{fig:schematic}e-f illustrates the weighting accuracy, which shows the resulted weights (red dots) of the two MRRs being examined at the tested values represented by each grid point. The gray dots correspond to the same targeted weight without dithering control. An improvement of over 2 bits (from 6.7 to 9 bits) is observed, enabling the MRR weight bank to have competitive performance with its electronic counterparts.

While our setup included the demonstration of the wireless transceiver system, we also had a versatile control setup that provided flexibility and accuracy in controlling the carrier frequencies and the mixing matrix. In the control setup, the signal mixtures were generated by a high-speed multi-channel arbitrary waveform generator (N8196A, 92GSPS, Keysight) and sent to each MZM directly. The generation of the two baseband signals, the up-conversion, and the signal mixing, is performed via software tools (Python). The mixed signals are then transmitted to the photonic system.

\section{Theoretical impact of weighting accuracy}

BSS must be able to deal with mixtures that are difficult to separate. Denoting the $j$th signal component in the $i$th mixture as $y_{ij}(t)$, the signal-to-interference ratio (SIR) \cite{choi2005blind} defined as

\begin{equation}
\textbf{SIR}_i(dB) = 10\mathrm{log}_{10}\frac{\|y_{ij'}(t)^2\|}{\Sigma_{j\neq j'}\|y_{ij}(t)^2\|}
\label{eq:SIR}
\end{equation}
is a ratio of the signal power ($\|y_{ij'}(t)^2\|$) to the rest interference power ($\Sigma_{j\neq j'}\|y_{ij}(t)^2\|$) of the $i$th mixture. Given a problem with $N$ mixtures containing $N$ sources to be separated, an often used merit is the overall SIR, which is the average of the SIR of every mixture ($\frac{1}{N}\Sigma_i \textbf{SIR}_i, i=1,2,\cdots N$). The SIR captures the accuracy of the system, and it can also be stated as a function of frequency, bandwidth, or other metrics. A higher SIR means better suppression of the interference signals. This metric of SIR can be extended to any mixing \textbf{H} through the ill-condition number, $\kappa (\mathbf{H})$ \cite{belsley2005regression}, defined as

\begin{equation}
\kappa (\mathbf{H}) = \mathrm{\|}\mathbf{H}\mathrm{\|}\cdot\mathrm{\|}\mathbf{H}^{-1}\mathrm{\|}.
\label{eq:ill_condition}
\end{equation}
This describes the demixing difficulty calculated by the mixing matrix \textbf{H}. Mixtures with a small ill-condition number are easier to solve. Conversely, problems with a sizeable ill-condition number are challenging and prone to smaller SIR. Typically, an ill-conditioned BSS problem requires the weighting to represent the inverse matrix accurately.

\begin{equation}
\textbf{H}^{-1} = \frac{a}{2a-1}
\begin{bmatrix}
1 & (a-1)/a\\
(a-1)/a & 1\\
\end{bmatrix}
\label{eq:H}
\end{equation}

To prove this, consider a simple case where the mixing matrix is symmetrical such that $\textbf{H} = [[a, 1-a],[1-a, a]], a\in[0.5, 1]$, which is often the case when two receiver antennas and the two transmitter sources are symmetrically positioned and have identical power. The inverse of matrix $\textbf{H}$ is given in Eq. \ref{eq:H}. To maximize the output signal amplitudes and introduce the weighting error, Eq. \ref{eq:H'} describes the actual matrix of mixtures that is input to the photonic BSS. This matrix is also scaled up for the maximal output weight, $w=1$. $\delta$ denotes the weight error caused by the inaccurate MRR control.

\begin{equation}
\textbf{H'}^{-1} = 
\begin{bmatrix}
1+\delta & (a-1)/a+\delta\\
(a-1)/a+\delta & 1+\delta\\
\end{bmatrix}
\label{eq:H'}
\end{equation}

The coefficients (proportional to the amplitudes of each source signal) of the separated results can then be expressed by Eq. \ref{eq:S}; the percentage of error (ignoring the noise) that remains in the recovered results can be represented as $a\delta/(2a+2a\delta-1)$. Thus, if $a=0.9$ as given in this problem, the control accuracy improvement from 6.7 bits ($\delta=2/2^{6.7}\approx0.019$) to 9 bits ($\delta\approx0.004$) could, in theory, lower the error by a factor of 4.6 (from $2.0\%$ to $0.44\%$). With these equations, we can also derive that if $2\%$ error ($\textbf{SIR}\approx33$) is demanded, this 2.3-bit improvement in control accuracy increases the solvable ill-condition number from 2.05 ($a=0.9$) to 10 ($a=0.55$).

\begin{equation}
\begin{split}
[s'_1, s'_2] & = \textbf{H'}^{-1}[r_1,r_2]^T = \textbf{H'}^{-1}\textbf{H}[s_1,s_2]^T\\
& = 
\begin{bmatrix}
(2a-1)/a+\delta & \delta\\
\delta & (2a-1)/a+\delta\\
\end{bmatrix}
\begin{bmatrix}
s_1\\
s_2\\
\end{bmatrix}
\label{eq:S}
\end{split}
\end{equation}

\section{Demonstration on wireless transceiver}

\begin{figure}[ht]
\centering
\includegraphics[width=.95\linewidth]{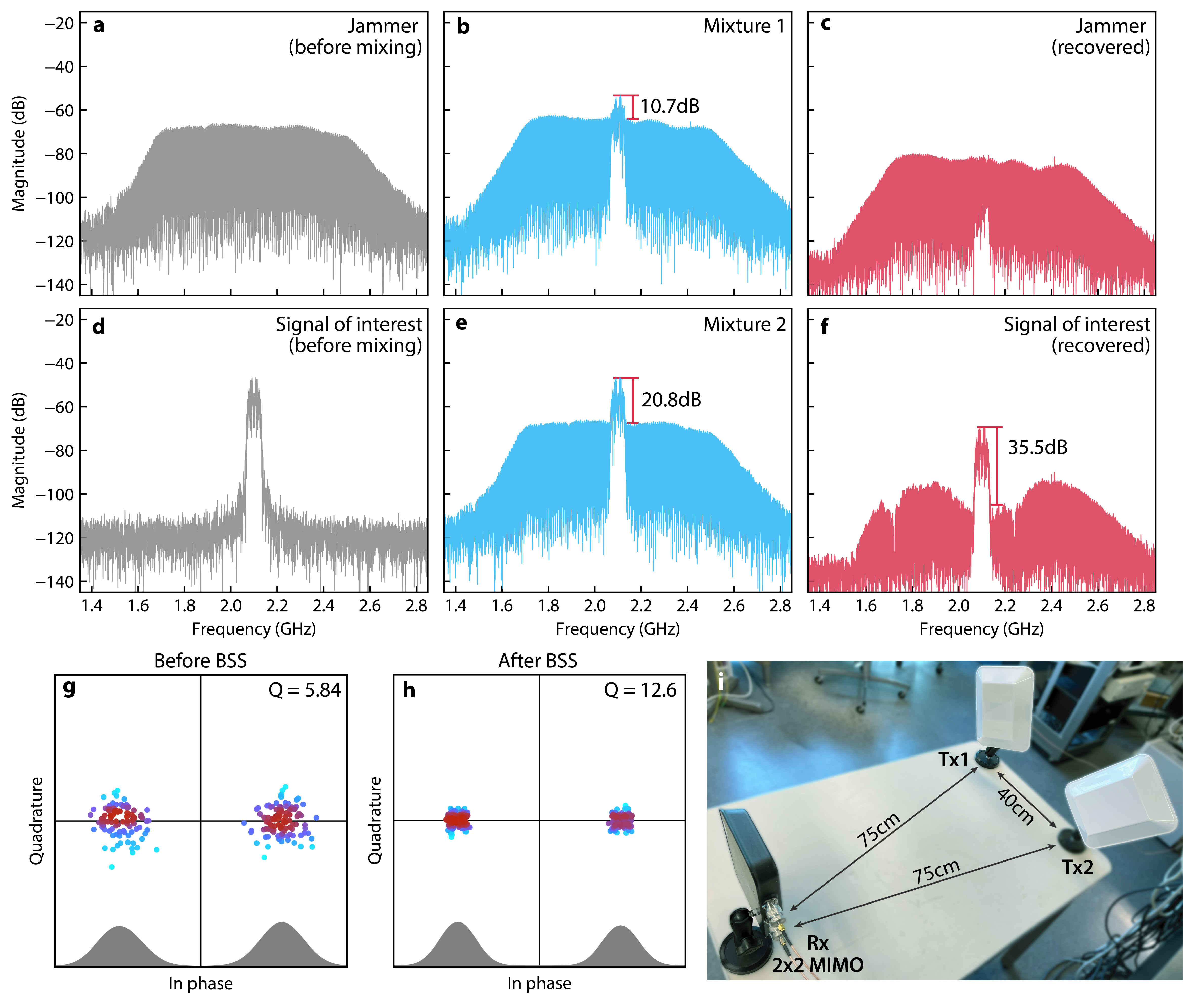}
\caption{\textbf{BSS demonstration on wireless transceiver system.} \textbf{a}, \textbf{d}, Spectrum of the source (Tx1) and the jammer (Tx2). These were measured at the receiver when one of the two transmitters (Tx1 or Tx2) was tuned off. \textbf{b}, \textbf{e}, Spectrum of received mixtures. \textbf{c}, Spectrum of the separated jamming noise. \textbf{f} Spectrum of the recovered source. \textbf{g},\textbf{h}, Constellation diagram of received mixture \textbf{g} (also corresponing to (\textbf{e})) and recovered source (\textbf{h}). The Q values are 1.90 and 5.84 for the mixtures corresponding to \textbf{b} and \textbf{e}, respectively. \textbf{i}, Antenna setup.
}
\label{fig:antenna_exp}
\end{figure}

Based on the setup described above, we firstly demonstrated our proposed BSS photonic processor in a wireless transceiver system, which emulates the case where a communication link is deteriorated by nearby RF interference. As shown in Fig. \ref{fig:antenna_exp}i, two antennas (1009-002, 1.7-2.5 GHz, Southwest Antennas) transmit the signal of interest and a broadband jammer mixed over the air with a transmission distance of 0.75 m. Then, the mixtures are received by a 2x2 MIMO antenna (1055-368, 1.7 - 2.5 GHz, Southwest Antennas), with two outputs corresponding to the polarization of 45-degree slant left and 45-degree slant right. The transmitted signal carries a repeating sequence of 200-random bits at a baud rate of 50 MHz with a binary phase-shift keying (BPSK) modulation format and a carrier frequency of 2.1 GHz. The interference is an instantaneous broadband jamming signal generated by adding 10,000 single tones with random phases and has a spectrum from 1.7 GHz to 2.5 GHz, covering the entire bandwidth of the antennas. Thus, extracting the signal of interest is insignificantly effective via spectral filtering even if the carrier frequency is known. If doable, the signal-to-noise ratio remained the same with the received mixtures at best. In contrast, photonic BSS can recover the communication link with no prior assumptions and suppress the interference noise. Fig. \ref{fig:antenna_exp}a-h illustrates the spectrum and the constellation diagram before and after the BSS process. The signal-to-noise ratio has almost 15 dB improvement (20.8 dB to 35.5 dB), accompanied by a 2-fold increase of the Q value (5.84 to 12.6) from the constellation. This result demonstrated effective suppression of the nearby interference with high power and broad bandwidth, maintaining the transmission quality of the wireless communication link.

\section{Broadband capability}

\begin{figure}[ht]
\centering
\includegraphics[width=.99\textwidth]{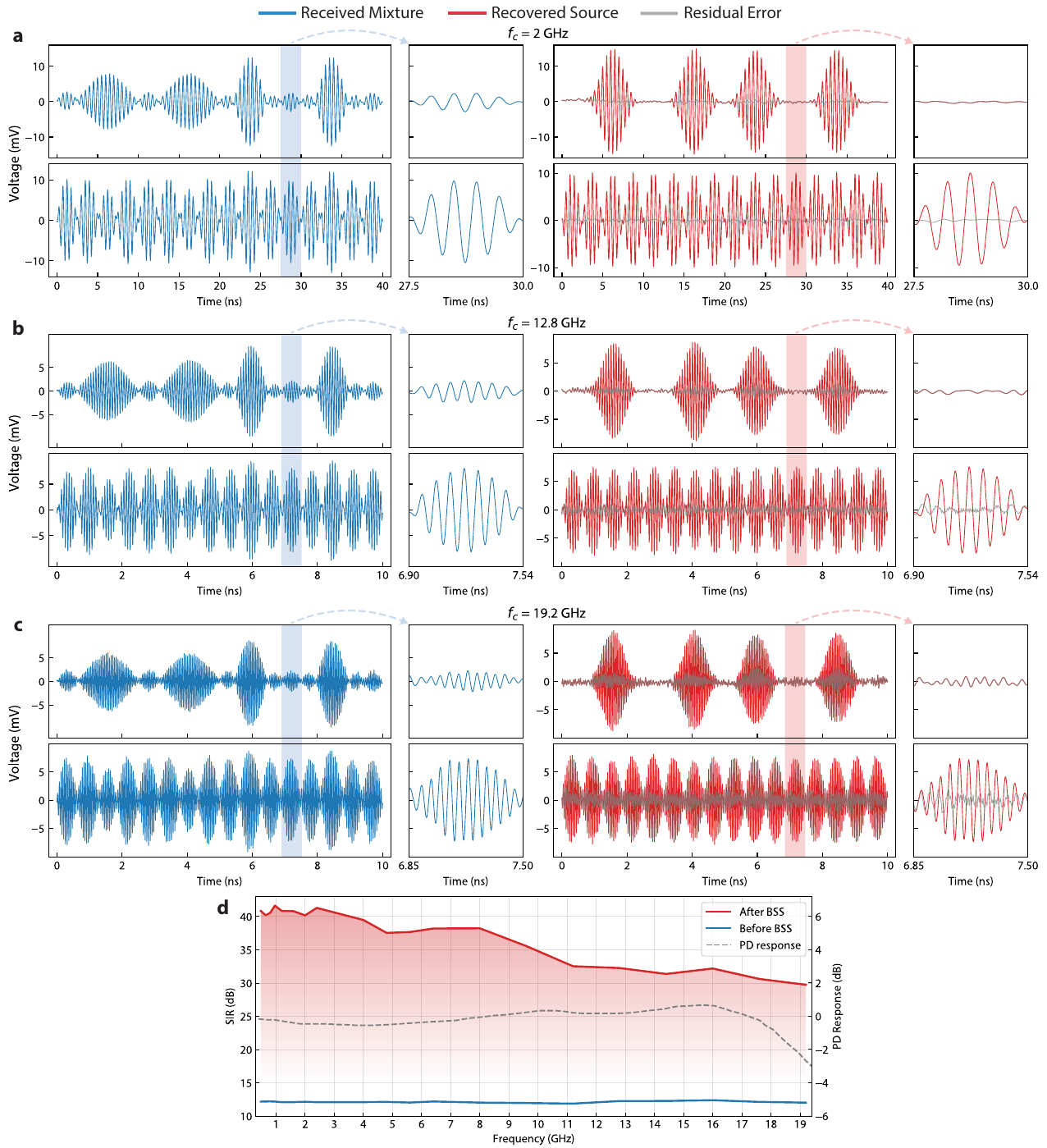}
\caption{\textbf{Broadband BSS with a single photonic chip.} \textbf{a}-\textbf{c}, BSS results, the mixtures and estimated sources with carrier frequencies of 2 GHz, 12.8 GHz, and 19.2 GHz, respectively. \textbf{d}, The SIR of the received mixtures (blue curve) and the recovered sources (red curve) versus the carrier frequency. The dashed gray curve is the relative power response of the photodetector.
}
\label{fig:bandwidth}
\end{figure}

Next, we examined the proposed photonic BSS system under different scenarios, with different signal mixtures enabled by programming the AWG. The original two signals are repeating patterns of 16 bits, and are in format of binary phase-shift keying (BPSK; $\textrm{bit pattern} = [0,1,0,1,0,1,0,1,0,1,0,1,0,1,0,1]$) and on-off keying (OOK; $\textrm{bit pattern} = [0,0,1,0,0,0,1,0,0,1,0,0,0,1,0,0]$), respectively. This configuration provided two short data periods that can be easily illustrated and also contributed all the potential bits combinations in their mixtures for a complete examination. Based on this setup, we firstly test the broadband capability by performing BSS on mixtures of signals from 1 GHz to 19.2 GHz by varying the carrier frequencies, as shown in Fig. \ref{fig:bandwidth}. The baseband frequencies were also adjusted according to the carrier frequencies, which were 160 MHz for less than 1 GHz, 400 MHz for 1 - 3 GHz, 800 MHz for 3 - 6 GHz, and 1600 MHz for $f_{\mathrm{carrier}} \geq 4.8 \mathrm{GHz}$. Annotating the two mixtures with $\mathrm{M1}$ and $\mathrm{M2}$ and the two sources with $\textrm{S1}$ and $\mathrm{S2}$, the mixing can be expressed as $\mathrm{M1} = 0.8\times\mathrm{S1} + 0.2\times\mathrm{S2}$ and $\mathrm{M1} = 0.2\times\mathrm{S1} + 0.8\times\mathrm{S2}$, denoting an ill-condition number of 2.26 (according to Eq. \ref{eq:ill_condition}).

As shown in Fig. \ref{fig:bandwidth}d, the 22 tested frequencies from 1 GHz to 19.2 GHz show SIR of no less than 30dB. Compared with previous photonic demonstration \cite{bss_20} that dealt with problems of a similar ill-condition number, we obtained almost 55 times broader bandwidth (19.2 GHz versus previously 350 MHz centered at 900 MHz) while maintaining a clean signal separation across the entire band (SIR $>$30 vs. previously SIR$\approx$14). This improvement in error suppression confirms the benefit of the improved dithering control method. Also, based on the Federal Communications Commission (FCC) frequency allocation chart (partly shown in Fig. \ref{fig:vision_figure}a), this broadband coverage by a single silicon photonic chip translates into the agility of processing multiple commonly used bands. Examples of included bands are cellular (620 MHz - 6.425 GHz), Wi-Fi (2.4 GHz, 5 - 7.125 GHz), military radar (extensive spectral usage above 8.5 GHz), and those for earth explorer satellite and radio astronomy (sparsely spread from 1.4 - 17.3 GHz). This wide bandwidth can also provide full coverage to some challenging bands, such as the ultra-wideband (UWB; 3.1 - 10.6 GHz). We also note that since signals remain narrowband at higher frequencies for photonic devices and the state-of-the-art BPD can be 100 GHz or more \cite{lischke2021ultra}, this system can easily expand the coverage to other important spectrums like millimeter-wave just by using higher bandwidth photodetectors. The practical device footprint on-chip is $0.13\,\mathrm{mm}\times0.42\,\mathrm{mm}$ including four MRRs and the waveguide routing, which slowly scales up in a linear fashion with the number of sources.

\section{Test on ill-conditioned mixing}

\begin{figure}[ht]
\centering
\includegraphics[width=.6\linewidth]{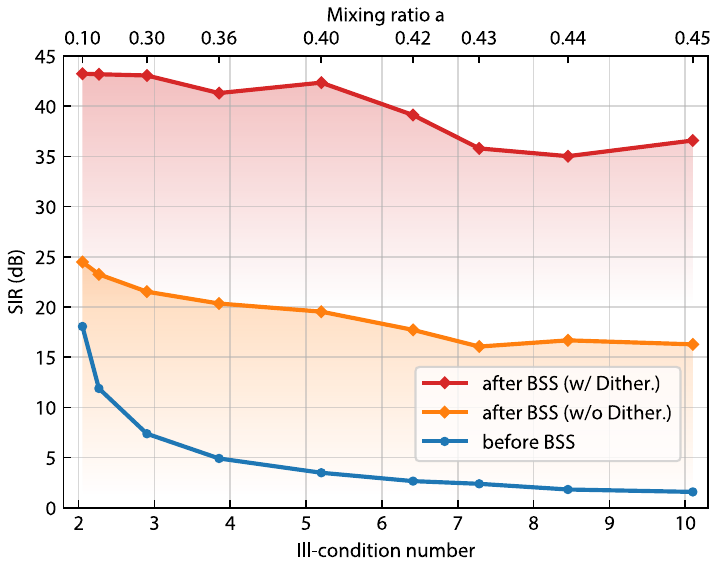}
\caption{\textbf{BSS performance on ill-conditioned mixtures.} Blue, red, and orange curves are the SIR before the BSS, after BSS with dithering control, and after BSS without dithering control, respectively. The mixing matrix $\textbf{H}$ is defined by the mixing ratio $a$ by $ \textbf{H} = [[a, 1-a],[1-a, a]]$. The corresponding ill-condition number is calculated by Eq. \ref{eq:ill_condition}.
}
\label{fig:ill_condition}
\end{figure}

Last, we investigated the performance of the proposed photonic BSS system in solving problems with different ill-condition numbers, which is to justify the significance of the improved weighting accuracy. Here, we fixed the carrier frequency at 1 GHz. The mixing matrix is of symmetrical form ($ \textbf{H} = [[a, 1-a],[1-a, a]]$) where $a$ is varied from 0.1 to 0.45, resulting in ill-condition numbers ranging from 2.05 to 10.1. Fig. \ref{fig:ill_condition} shows the SIR of the mixtures obtained from the same setup but with the dithering control (compared to the previous control method without the dithering). Even in the presence of similar experimental noise levels, lower SIR is always obtained when not using dithering control. Conversely, the dithering controlled setup maintained constantly high SIR such that the influence of ill-conditioning was less distinguishable. The average SIR is above 35 dB for all tested cases, which generally shows around 20 dB improvement compared to the previous control method (the orange curve in Fig. \ref{fig:ill_condition}), as expected in the analysis part above. This improvement confirms the significance of accurate weight control for MRR-based applications, such as the BSS in this paper.

\section{Conclusion}

In summary, we explored a physical layer cognitive radio solution based on BSS being performed on a dithering controlled silicon MRR weight bank. This solution is a complete RF frontend with a photonic signal processor that can do intelligent learning through the fully programmable and integrated electronic-photonic system. This setup has an unprecedentedly high bandwidth of up to 19.2 GHz that can fully demonstrate the capability of the BSS technique. In addition, the high SIR observed for all the frequencies and ill-conditioned problems, together with an example of a wireless transceiver system, confirms the benefits of real-world applications brought by the improved MRR control method. The superior performance of this photonic approach illustrates the readiness to replace conventional RF implementations, effectively addressing the incoming challenges in wireless communications, including bandwidth limitations, energy efficiency, and latency. 

With the availability of higher speed modulators \cite{he2019high} and photodetectors \cite{lischke2021ultra} on the silicon platform, as well as the maturity of packaging, including photonic wire bonding \cite{lindenmann2012photonic}, laser integration \cite{zhou2015chip}, co-packaging of silicon and CMOS chips, and monolithic cointegration of CMOS and silicon photonics \cite{zerochange}, this proposed photonic BSS can have future implementations with higher integration and broader bandwidth. With these, we envision a standalone BSS device of a small form factor that is field-deployable for various applications, including interference cancellation in autonomous vehicles and aviation.

\section{Methods}
%Name something might be worth include here
\subsection{Photonic BSS algorithm}
The photonic BSS algorithm follows a pipeline including principal component analysis (PCA), whitening, and independent component analysis (ICA). Firstly, photonic PCA updates the MRR weights to converge at the target PC vectors by maximizing the variance of weighted addition output $E(y^2_i)$. The obtained PC vectors and variances can construct a whitening matrix, which in the following step can reduce the BSS problem to finding an orthogonal transformation (rather than the inverse matrix of a mixing matrix). Then, photonic ICA utilizes this whitening matrix $V$ and updates the MRR weight vector in the whitened subspace $w_i V(i = 1,\dots, n)$ to converge at the target IC vectors by maximizing the absolute relative kurtosis of the output, calculated as $\mid E(y^4_i )/\sigma^4(y_i)-3\mid$ ($\sigma^2$ is variance). This algorithm is inspired by the central limit theorem that the sources are supposed to maximize their non-Gaussianity (relative distance from the Gaussian distribution). Our BSS pipeline does not require explicit waveform digitization but only the second-order and fourth-order statistics of weighted addition output, unlike conventional BSS solutions like FastICA. This feature permits a low-cost ADC and DSP working at the sub-Nyquist sampling regime, generally more applicable to broadband BSS.

\subsection{Fully packaged photonic processor}
We fully packaged the silicon photonic chip with its driver and control circuitry in the same interposer PCB in our experimental setup, as shown in Fig. \ref{fig:schematic}d. This integration benefits a simplified lab setup, lower power consumption (<100 mW), and neat connectivity that a single USB cable handles both the digital interface and the power delivery. In terms of power management, the 5 V input (from USB cable) generates isolated and low-noise power rails by several dedicated power management ICs (LT1533CS and two LT3042, Analog Devices). These rails power up the precision current sources (LTC2662-16, 16-bit, Analog Devices), which drive the MRRs on the photonic chip. The digital interface between the current source and the controller (Arduino Pro Micro, Sparkfun) is built on the SPI protocol and isolated by a dedicated IC (ADUM4151, Analog device). This fully isolated MRR driving circuity helps with noise suppression. Host PC commands the weights through serial communication via the USB cable. The onboard controller phrases the received command and talks to the current sources to adjust the current of each microring.

\subsection{Device fabrication}
The silicon photonic chip was fabricated on a silicon-on-insulator wafer with a silicon thickness of 220 nm and a buried oxide thickness of 2 µm. The waveguide is 500 nm wide. The weight bank consists of four MRRs (radius around $\textrm{r}=22\;\upmu\textrm{m}$) coupled with two bus waveguides in an add/drop configuration, and two among the four were used in the experiments. A slight difference ($\Delta\textrm{r}=0.32\;\upmu\textrm{m}$) was introduced in the ring radii to avoid resonance collision. The gap between the ring and bus waveguide is 200 nm, yielding a Q factor of about 6,000. Circular metal heaters were built on top of each MRR for thermally weight tuning. Metal vias and traces were deposited to connect the heater contacts of the MRR weight bank to electrical metal pads.

\section{Data Availability}
All data used in this study are available from the corresponding authors upon reasonable request

\section{Code Availability}
All codes used in this study are available from the corresponding authors upon reasonable request.

\backmatter

% \bmhead{Supplementary information}

% \begin{appendices}

% \section{Section title of first appendix}\label{secA1}

% \end{appendices}


\bibliography{ref}

\section{Acknowledgments}

This research is supported by the National Science Foundation (NSF) (ECCS-2128608 and ECCS-1642962), the Office of Naval Research (ONR) (N00014-18-1-2297 and N00014-20-1-2664), and the Defense Advanced Research Projects Agency (HR00111990049). The devices were fabricated at the Advanced Micro Foundry (AMF) in Singapore through the support of CMC Microsystems. B. J. Shastri acknowledges support from the Natural Sciences and Engineering Research Council of Canada (NSERC). S. Bilodeau acknowledges funding from the Fonds de recherche du Québec - Nature et technologies.

\section{Author contributions}
W.Z., T.F.L. and A.T. conceived the ideas and implemented the experimental setup, designed the experiment, conducted the experimental measurements and analysed the results. T.F.L., E.C.B., S.B. and A.J. designed the silicon photonic chip. T.F.L, C.H., A.T. and B.J.S. provided the theoretical support. E.C.B. and S.B. performed the chip packaging. W.Z., A.T., C.H., S.B. and B.J.S. wrote the manuscript. P.R.P. supervised the research and contributed to the general concept and interpretation of the results. All the authors discussed the data and contributed to the manuscript.

\section{Competing interests}
The authors declare no competing interests.

\end{document}